\documentclass[conference]{IEEEtran}
\usepackage{amsmath,amssymb,amsfonts}
\usepackage{algorithmic}
\usepackage{graphicx}
\usepackage{textcomp}
\usepackage{xcolor}
\usepackage{lipsum}
\usepackage{cite}
\usepackage{breqn}
\usepackage{pifont}
\newcommand{\cmark}{\ding{51}}%
\newcommand{\xmark}{\ding{55}}%

\usepackage[T1]{fontenc}    
\usepackage{amsfonts}       
\usepackage{amsmath}        
\usepackage{amssymb}        
\usepackage{booktabs}       
\usepackage{caption}        
\usepackage{subcaption}
\usepackage{float}          
\usepackage[hidelinks]{hyperref}       
\usepackage{microtype}      
\usepackage{rotating}       
\usepackage{tikz}           
\usepackage{url}            
\usetikzlibrary{arrows,decorations.pathmorphing}
\usepackage[ruled,vlined,linesnumbered]{algorithm2e}
\usepackage{ragged2e}
\usepackage{booktabs}
\usepackage{siunitx}
\usepackage{pgfgantt}
\usepackage[arrowdel]{physics}

\def\BibTeX{{\rm B\kern-.05em{\sc i\kern-.025em b}\kern-.08em
    T\kern-.1667em\lower.7ex\hbox{E}\kern-.125emX}}
    
\begin{document}

\title{DVGAN: Stabilize Wasserstein GAN training for time-domain Gravitational Wave physics}

\author{\IEEEauthorblockN{Tom Dooney}
\IEEEauthorblockA{\textit{Centre of Actionable Research} \\
\textit{Open University}\\
Heerlen, Netherlands \\
td1@ou.nl}
\and
\IEEEauthorblockN{Stefano Bromuri}
\IEEEauthorblockA{\textit{Centre of Actionable Research} \\
\textit{Open University}\\
Heerlen, Netherlands \\
stefano.bromuri@ou.nl}
\and
\IEEEauthorblockN{Lyana Curier}
\IEEEauthorblockA{\textit{Centre of Actionable Research} \\
\textit{Open University}\\
Heerlen, Netherlands \\
lyana.curier@ou.nl}

}

\maketitle

\begin{abstract}

Simulating time-domain observations of gravitational wave (GW) detector environments will allow for a better understanding of GW sources, augment datasets for GW signal detection and help in characterizing the noise of the detectors, leading to better physics. This paper presents a novel approach to simulating fixed-length time-domain signals using a three-player Wasserstein Generative Adversarial Network (WGAN), called DVGAN, that includes an auxiliary discriminator that discriminates on the derivatives of input signals. An ablation study is used to compare the effects of including adversarial feedback from an auxiliary derivative discriminator with a vanilla two-player WGAN. We show that discriminating on derivatives can stabilize the learning of GAN components on 1D continuous signals during their training phase. This results in smoother generated signals that are less distinguishable from real samples and better capture the distributions of the training data. DVGAN is also used to simulate real transient noise events captured in the advanced LIGO GW detector.
\end{abstract}


\section{Introduction}\label{Introduction}

Generating high-fidelity, synthetic time-domain signals has applications in many areas including medical \cite{medicine_GAN}, music \cite{music_generation_survey}, and particularly relevant to this study, gravitational wave (GW) physics \cite{first_GW}. GW physics has entered the forefront of astrophysics and cosmology since the first detection of gravitational waves in 2015, setting the course for a new era of observational astrophysics. Since then, many other black hole mergers have been detected by advanced LIGO and advanced Virgo detectors \cite{CBC_detections_1}\cite{CBC_detections_3}. 

Compact binary coalescence mergers, such as black hole mergers, are detected through the process of matched-filtering \cite{Owen_1999_match}, where detector outputs are compared against a large template bank of modelled waveforms that are calculated from Einstein's General Relativity equations. GW physics also requires the identification and study of unmodelled, transient waveforms embedded in detector data, known as GW `bursts'. Bursts are typically of short duration, and cannot be accurately modelled with current technology. Astrophysical sources for burst events include events such as core-collapse supernovae \cite{CCSNa} and neutron star post-mergers \cite{Neutron_star}, although other potential sources have yet to be explained. Unlike modelled searches, burst waveforms cannot be identified by matched-filtering. Instead, such model-free detection algorithms typically investigate excess power in the time-frequency representation of the data to identify the signal from noise \cite{wave_burst}, and rely on such signals being picked up in multiple detectors. These methods are hindered by the occurrence of non-Gaussian transient noise events, known as 
‘glitches'. Glitches have durations typically on the order of sub-seconds and can be difficult to differentiate from signals originating from astrophysical sources. Their causes can be environmental (e.g., earthquakes, wind) or systematic (e.g. scattered light), although in multiple cases, their sources remain unknown \cite{detchar_transient}.

Ongoing upgrades to advanced detector systems, and in particular, the introduction of next-gen GW detectors in the coming years, such as the Einstein Telescope (ET) \cite{ET_paper} and Cosmic Explorer \cite{Cosmic_expl}, will give rise to new challenges in GW data analysis. Modelled and unmodelled detections originating from all sources are expected to significantly increase i.e. it is estimated that the ET will detect one compact binary coalescence (CBC) event approximately every $6s$. Furthermore, there will be many classes of sources visible in the detectors at the same time, while other sources interpreted as transients in current detectors will occur in next-gen detector bands for hours or even days (i.e. inspiral signals could stay within sensitive frequency bands for up to 10 days, for the lightest systems, and as briefly as a few 100 milliseconds for the heaviest ones), resulting in overlapping signals. This will lead to higher uncertainty in parameter estimations of GW sources. 

Perhaps the biggest challenge will lie in matched-filter analysis since the number of templates grows approx. as $f_s^{-11/3}$, where $f_s$ is the frequency below which a negligible amount of signal-to-noise ratio is accumulated. The value of $f_s$ could be a factor of 20-40 times smaller for next-gen detectors, resulting in a massive increase in search templates and alarm rates, possibly rendering matched-filtering techniques infeasible\cite{ET_paper}. These environments will require a paradigm shift in the way GW data is currently analyzed, taking GW physics into the realm of Big Data. New efficient, online reconstruction methods will be required to handle the vast quantities of data from multiple detector outputs and the expanded signal space.

Although still in its infancy in GW physics, deep learning (DL) has shown great potential in GW data analysis, matching the sensitivity of match-filtering for advanced LIGO and advanced Virgo GW searches \cite{match_matched}. In cases where waveforms of interest are rare, are unmodelled, or are not well understood, datasets of sufficient size cannot be constructed to train DL models. When real data is limited, the ability to simulate representative synthetic data is particularly valuable. Generative Adversarial Networks (GANs) \cite{GANpaper} have grown popular in synthetic data generation since their inception in 2015. Although their impact has been largely seen in computer vision, recently their utility has been increasingly realised in time series and sequence generation \cite{TimeGANsSurvey}. In this paper, we employ a new variant of 1D GAN, called Dual-Discriminator Derivative GAN (DVGAN) that includes two discriminators rather than the usual single discriminator to simulate proxy GW signals. A second auxiliary network is trained to distinguish the derivatives of real and synthetic signals during training. 

This paper is structured as follows; in section \ref{Related_work}, we address the state-of-the-art (SOTA) in synthetic sequence generation and current data augmentation approaches in GW physics, and motivate the data augmentation framework proposed in this study. In section \ref{Concepts}, we discuss concepts relating to GANs and their application to sequential data. In section \ref{Methods}, we present the methods followed in this research, introducing the datasets, the DVGAN architecture and the training scheme. In section \ref{Results}, we present the results of synthetic data produced by the DVGAN under benchmark experiments, comparing them with the vanilla two-player case that otherwise has the same base architecture. Finally, section \ref{Conclusion} presents the conclusions of this research with avenues for future research.

\begin{figure}[t!]
\centering
\captionsetup{width = 0.45 \textwidth}
  \includegraphics[width=0.43\textwidth]{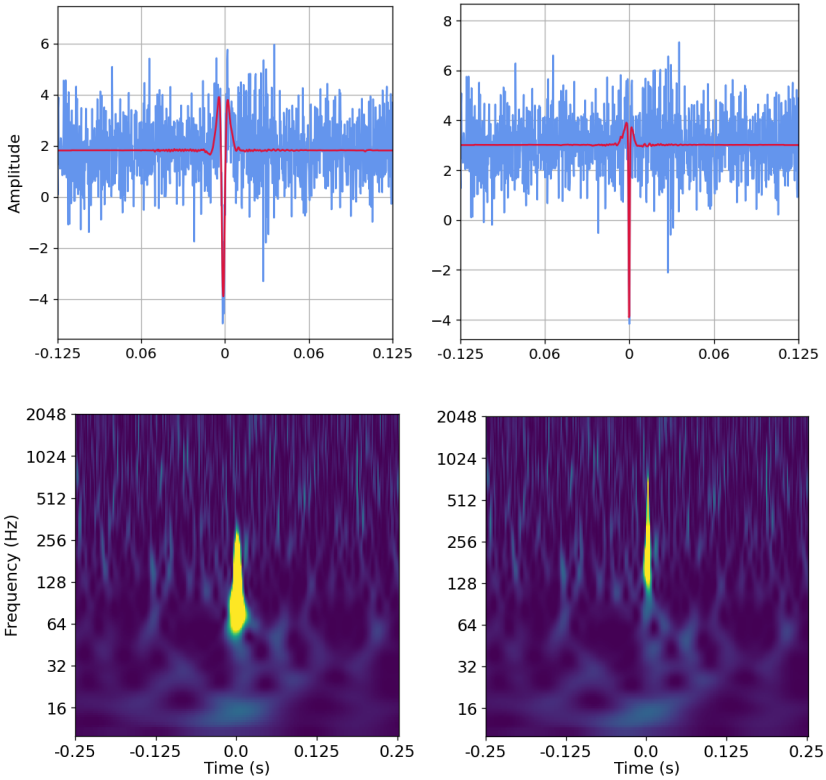}
\caption{Time series (top) and corresponding Q-transform (bottom) representations of blip glitches from LIGO's Hanford (H1) detector\cite{GENGLI}.}
\label{fig:TS_Spec}
\end{figure}


\section{Related Work}\label{Related_work}
Most of the attention brought to GANs \cite{GANpaper} has remained in computer vision and image generation. In particular, they have been studied extensively in the context of GW physics to augment datasets where studies have largely been restricted to the case of 2D spectrogram data\cite{GAN_spec_best}. This is not surprising, since important algorithms used to identify GW waveforms utilize spectrogram representations of time-series signals (eg. \cite{Omicron}). However, a relatively new niche for GANs has emerged in the development of high-fidelity and diverse time series data.

There are multiple reasons why simulating the time-series representation of GW strain data is desirable. Firstly, this is the representation of the data that is directly measured by GW detectors, and if DL algorithms can be trained in the 1D space, this may circumvent the need for computationally expensive transformations to the time-frequency representation. Second is the flexibility offered by simulating time-domain signals, where signals can be easily manipulated, in contrast to spectrogram data. This might provide an avenue to explore the space of overlapping signals, a problem expected from next-gen detectors, as described in Section \ref{Introduction}. Finally, phase information is lost through spectrogram transforms. Should relevant waveforms be simulated effectively, transforming to the spectrogram domain is trivial, while the reverse is not.

The work presented in this paper falls in the realm of continuous time series data generation. Multiple studies have implemented the GAN framework within the temporal setting, many of which are described in \cite{TimeGANsSurvey}. These works typically inherit recurrent-based networks (i.e. LSTM, GRU) within the GAN components \cite{RCGANpaper}\cite{TimeGANpaper}.

A prominent challenge in time series data generation is that long time series data streams can greatly increase the dimensionality requirements of generative modelling using recurrent-based architectures. This may render such approaches infeasible for the datasets applicable to this research, which comprise large sequences of hundreds and even thousands of data points. In the previously described studies, they have proven their effectiveness on relatively short time series signals. One study attempts to address this issue by developing a metric called Sig-Wasserstein-1 that captures time series models’ temporal dependency and uses it as a discriminator in a time series GAN \cite{sigWasserstein}. However, its performance has only been reported on non-differentiable financial datasets rather than smooth waveforms.

One study that makes progress in generating longer sequences analogous to GW waveforms is presented in \cite{McGinn_2021}, where a GAN framework is implemented on generalised burst waveforms that are analogous signals to those observed in GW detectors. They condition a convolutional GAN, called McGANn, on 5 different waveform classes and show that they can generate class-interpolated, hybrid waveforms by conditioning their GAN on a mix of the input classes through the continuous sampling of the class space. They show the usefulness of their generated data by training CNNs on detecting the waveforms in additive Gaussian noise, training them on waveforms generated by different sampling methods in their class space.

A similar approach to simulating noise events is presented in \cite{GENGLI}\cite{GENGLI_2}. In this study, a 1D convolutional Wasserstein GAN is implemented on actual glitch events found in the Advanced LIGO detector. The study deals with one glitch class, the blip glitch. Their work relies on that of \textit{Gravity Spy} \cite{Zevin_2017}, a deep learning classifier used to classify gravitational wave and glitch events that uses citizen science and domain expert knowledge to label training data observed in GW detectors \cite{gspy_data_quality}. They evaluate their synthetic data using confidences provided by the \textit{Gravity Spy} classifier, while also using similarity metrics to evaluate their synthetic generations compared to the original training data. 

In contrast to the above studies, the work presented here aims to generate comparatively large 1D continuous signals using a novel GAN architecture comprising a three-player game with two discriminators and one generator, all being convolutional (with fully connected layers). GANs consisting of three players have already been explored for various tasks. For example, one study couples GANs with auxiliary classifiers for downstream tasks, encouraging the generation of samples that are difficult to classify for the classifier, thus boosting its performance \cite{3playerGANhard}. Another approach also uses an auxiliary classifier to perform facial recognition on real and generated samples from the GAN to encourage the generator to synthesize faces that preserve identity, allowing faster convergence \cite{3playerface}. 

One dual discriminator GAN study gives two discriminators the objective of minimizing both the KL and reverse KL divergences between data distribution and the distribution induced from the data generated by the generator, respectively \cite{3playerdual}. The authors show that this is an effective way to mitigate mode collapse (where only a few high-fidelity samples can be generated with little variety). While the three-player GAN presented in this work also aims to give additional adversarial feedback, it does this by looking at two different representations of the data, one being the original signal, and the other being the derivative signal. To the best of our knowledge, this approach has yet to be explored.

\section{Generative Adversarial Networks}\label{Concepts}
Typically as an unsupervised algorithm, GANs \cite{GANpaper} learn to generate representative samples of the training set through the adversarial process, in which two model components are set to compete with one another; a discriminator, which is designed to distinguish between real and fake data, with the objective of minimizing the classification error; and a generator, which performs a mapping from a fixed length vector $z$ to its representation of the data, with the objective of maximizing the discriminator's classification error on synthetic samples. 

The generator's input vector is drawn randomly from a Gaussian distribution, which is referred to as a latent space comprised of latent variables. The latent space is a compressed representation of a data distribution to which the generator applies meaning during training. Sampling points in this space allows the generator to produce a variety of different generations, with different points corresponding to different features in the generations. The discriminator maps its input $x$ to a probability that the input comes from either the training data (real) or the generator (fake). During training, the discriminator and generator are updated using batches of data. Random latent vectors are given to the generator to produce a batch of fake samples and an equal batch of real samples is taken from the training data. The discriminator makes predictions on the real and fake samples and the model is updated by minimising a loss function. 

\begin{figure*}[h]
\centering
\captionsetup{width = 0.85 \textwidth}
\includegraphics[width = 0.98\textwidth]{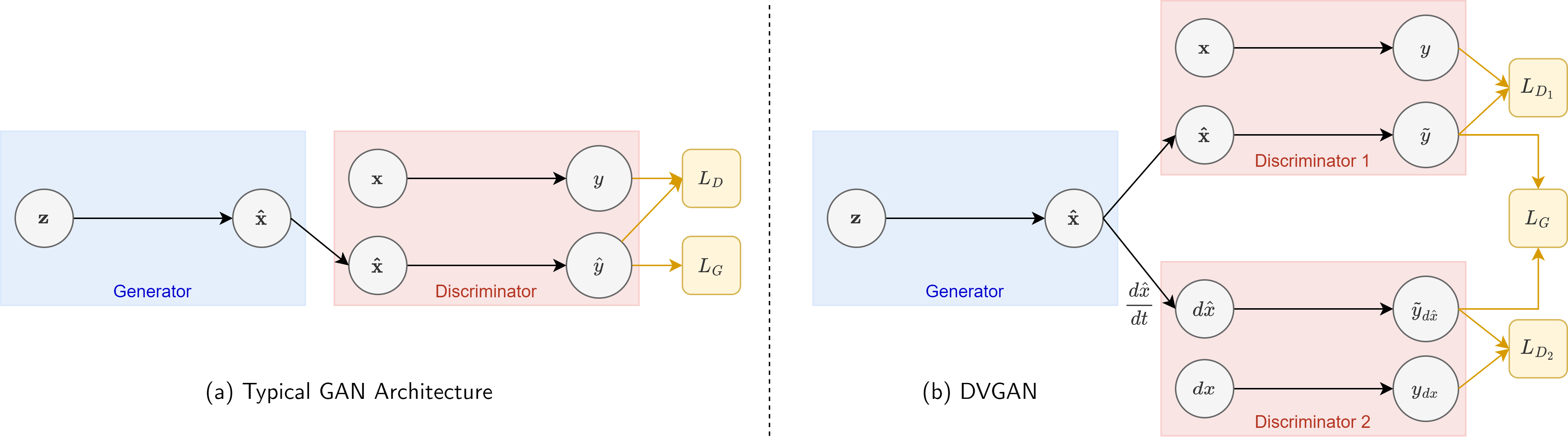}
\caption{Diagrams of a typical GAN architecture (\textit{left}) and DVGAN (\textit{right}).}
\label{fig:dual_discriminator_gan}
\end{figure*}

Early GAN approaches have been shown to work well under some problem conditions and model configurations, however, they suffer from issues such as vanishing gradients. GANs are known for their difficulty to train, and numerous studies have been centred around methods to stabilize the training process. Wasserstein GANs \cite{wGAN_paper} are a particular variant that use the Wasserstein-1 distance ($W1$) as the loss function, which measures the similarity between two distributions. $W1$ increases monotonically while never saturating, and is a meaningful loss metric, even for two disjoint distributions. Since it is continuous and differentiable, it yields reliable gradients, allowing for the discriminator to be trained until optimality. Under this paradigm, the optimization problem can be formulated as 

\begin{equation}\label{eq_wGAN_opt}
    \theta_{opt} = arg \min_{\theta}W1(P_x||P_{\hat{x}})
\end{equation}

where $P_x$ and $P_{\hat{x}}$ are real and generated distributions respectively. Equation \ref{eq_wGAN_opt} can be written as 

\begin{equation}
    \theta_{opt} = arg \min_{\theta}\max_{\phi: ||D(x, \phi)||_{L \leq 1}}L(\phi, \theta)
\end{equation}

with the discriminator loss given by

\begin{equation}\label{Discriminator_loss}
    L(\phi, \theta) = -\mathbb{E}_{x \sim P_x}[D(x, \phi)] + \mathbb{E}_{\hat{x} \sim P_{\hat{x}}}[D(\hat{x}, \phi)]
\end{equation}

where $\hat{x} = G(z, \theta)$ and $z$ being a batch of the generator's latent vector. $D$ and $G$ refer to the discriminator and the generator with parameters $\phi$ and $\theta$, respectively. $\mathbb{E}_{x \sim P_x}$ averages over a batch of real samples $x$, while $\mathbb{E}_{\hat{x} \sim P_{\hat{x}}}$ averages over a batch of generated samples $\hat{x}$. Equation \ref{eq_wGAN_opt} imposes the constraint of 1-Lipschitz continuity on $D$\cite{wGAN_paper}. Although this can be accomplished by clipping the weights of the discriminator, it has been shown that adding a regularization penalty called the gradient penalty ($GP$) to the discriminator loss is superior \cite{wGAN_GP_paper}. The discriminator loss then becomes 

\begin{equation} \label{eq_wGAN_gp}
    L_D = L(\phi, \theta) + \lambda GP(\phi)
\end{equation}

with 

\begin{equation}\label{eq_GP}
    GP(\phi) = \mathbb{E}_{\hat{x} \sim P_{\hat{x}}} \left[(||\nabla{x} D(\hat{x}, \phi)||_2 - 1 )^2\right]
\end{equation}

and where $\lambda$ represents the regularization hyperparameter, $||.||_2$
represents the L2-norm and $\hat{x}$ is evaluated following

\begin{equation}
    \hat{x} = \Tilde{x}t + x(1-t)
\end{equation}

with $t$ uniformly sampled $\sim [0, 1]$. 

This approach has been utilized in advanced GAN studies, such as \cite{karras2018progressive}. Also to note, this approach is not restricted to Wasserstein GANs \cite{improved_GAN_training}. While weight clipping can enforce the Lipschitz condition everywhere, the gradient penalty method does not. Wei et al. \cite{improving_improved_training} propose an additional regularization method to overcome this challenge where an additional penalization term is added to the loss from Equation \ref{eq_wGAN_gp}, known as the consistency term. This approach involves using dropout layers in the discriminator to achieve two perturbed versions of the real samples $x$. This results in two different estimates $D(x')$ and $D(x'')$, with a loss being calculated from the two perturbed outputs. This approach is omitted from this study for simplicity and to save computational expense, although it is considered that the proposed DVGAN method will be similarly effective under such regularization.

When updating the generator, errors are propagated through the entire network, from $D$ to $G$. Naturally, updates are made only on generated samples from $G$. The generator loss is written as

\begin{equation}
    L_G(\phi, \theta) = -\mathbb{E}_{\hat{x} \sim P_{\hat{x}}}[D(\hat{x}, \phi)]
\end{equation}

In the case of two discriminators being applied to two different representations of the input data, the total generator loss is written as

\begin{multline}\label{eq:combined_generator}
     L_G(\phi_1, \phi_2, \theta) = -\eta_1\mathbb{E}_{\hat{x}_1 \sim P_{\hat{x}_1}}[D_1(\hat{x}_1, \phi_1)]\\-\eta_2\mathbb{E}_{\hat{x}_2 \sim P_{\hat{x}_2}}[D_2(\hat{x}_2, \phi_2)]
\end{multline}

where $D_1$ and $D_2$, represent the first and second discriminators. Here $\hat{x}_1$ and $\hat{x}_2$ represent synthetic samples from both representations of the data,  $P_{\hat{x}_1}$ and $P_{\hat{x}_2}$ are the distributions of the two representations of generated data and $\eta_1$ and $\eta_2$ are hyperparameters to tune that control the relative strength of the losses from respective discriminators.

\section{Methods}\label{Methods}

\subsection{Dual-Discriminator Derivative GAN (DVGAN)} \label{DGAN_method}
This research is centred around a novel GAN architecture for generating 1D continuous signals. The adversarial process that is featured in GANs, where two model components compete against one another can result in a difficult and unstable training process. When dealing with complicated signals, GANs lacking the appropriate architecture can struggle to converge or can fall into a local minimum and suffer mode collapse. Finding equilibrium between the two components is the main challenge in training GANs. Ways to diagnose failures in convergence include using loss plots to identify failure modes and convergence. Motivated by the challenge of a convergent training process and noise artifacts commonly observed in 1D signals generated by WGANs, a second discriminator is also implemented that discriminates on the derivatives of real and fake signals, which is the primary novelty in this research. This additional model component can be seen as a further degree of regularization by giving information on the rate of change of signals. 

For the derivative discriminator, Equation \ref{Discriminator_loss} can be written as 

\begin{multline}\label{Discriminator_loss_DV}
    L(\phi_2, \theta) = -\mathbb{E}_{\frac{dx}{dt} \sim P_{\frac{dx}{dt}}}\left[D_2(\frac{dx}{dt}, \phi_2)\right]\\ + \mathbb{E}_{\frac{d\hat{x}}{dt} \sim P_{\frac{d\hat{x}}{dt}}}\left[D_2(\frac{d\hat{x}}{dt}, \phi_2)\right]
\end{multline}

while the total generator loss from Equation \ref{eq:combined_generator} combines the losses from both discriminators, written as

\begin{multline} \label{eq:combined_generator_DV}
     L_G(\phi_1, \phi_2, \theta) = -\eta_1\mathbb{E}_{\hat{x} \sim P_{\hat{x}}}\left[D_1(\hat{x}, \phi_1)\right]\\-\eta_2\mathbb{E}_{\frac{d\hat{x}}{dt} \sim P_{\frac{d\hat{x}}{dt}}}\left[D_2(\frac{d\hat{x}}{dt}, \phi_2)\right]
\end{multline}

where $D_1$ and $D_2$ represent the respective discriminators with weights $\phi_1$ and $\phi_2$.

The network architecture utilizes deep convolutional networks for the generator and both discriminators. GANs are sensitive to the size of components and their architectures, therefore, the discriminator and the generator of the WGAN maintain a similar number of parameters (4.1M), and mirror each other (almost exactly) in terms of their architecture. The derivative discriminator of DVGAN comprises a significantly smaller architecture with only about 8\% (300k) of the number of parameters of the base discriminator (while the base discriminator and generator are identical to the corresponding components of the WGAN). The generator uses strided transposed convolutions with BatchNormalization in the first layer and a stride of 2 for upsampling. 

\begin{figure}[t!]
\centering
\captionsetup{width = 0.45 \textwidth}
\begin{subfigure}{0.25\textwidth}
  \centering
  \includegraphics[width = \textwidth]{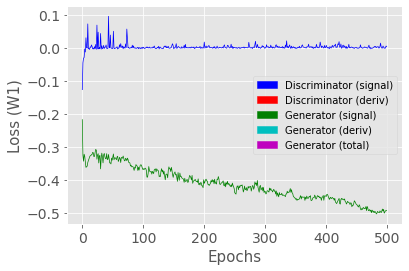}
  \caption{WGAN}
  \label{fig:Unstable_vanilla}
\end{subfigure}%
\begin{subfigure}{0.25\textwidth}
  \centering
  \includegraphics[width = \textwidth]{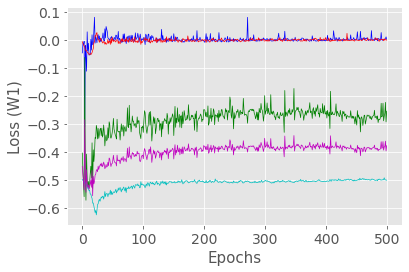}
  \caption{DVGAN}
  \label{fig:Stable_DGAN}
\end{subfigure}%
\caption{Examples of an unstable learning process from a vanilla WGAN (left) and a stable learning process from a DVGAN (right) on the Binary Black Hole (BBH) dataset.}
\label{fig:GAN_loss_comparison}
\end{figure}

The base discriminator mirrors the generator architecture without BatchNormalization but uses dropout and a stride of 2 in the convolutional layers for downsampling. Both discriminators and the generator employ a kernel size of 5 and use LeakyReLU and ReLU activations respectively, except for their last layers. A linear activation is used for the final generator layer (guaranteeing positive and negative outputs), and sigmoid activations are used for the final layer of both discriminators (representing the probability of a signal being synthetic). It is found that a smaller network can be utilized in the derivative discriminator and still yield superior results, with the focus of regularizing the training of the generator. A Wasserstein loss function is used with RMSprop as the optimizer, with a learning rate of $10^{-4}$. WGAN training can become unstable using momentum-based optimizers such as ADAM, and it is reported in the original WGAN paper that RMSprop provides more stable training \cite{Original_wasserstein}. The $GP$ regularization method described in Section \ref{Concepts} is also included in the losses of both discriminators. Throughout experimentation, $\eta_1$ and $\eta_2$ in Equation \ref{eq:combined_generator_DV} are both maintained at 0.5 (arbitrarily, optimization is not the focus of this study). Figure \ref{fig:GAN_loss_comparison} shows an example of the stabilizing effect the derivative discriminator has on the generator loss, compared to that of a vanilla WGAN. Unstable loss trends commonly encountered when training WGANs are one of the motivating factors of this research.

When training a GAN, the generator and discriminator must be updated at particular rates to achieve convergence. Updating the discriminator is more challenging since samples from the generator can be anywhere in the signal space and can change over each iteration \cite{discriminator_training}. Therefore, the discriminator is updated 5 times for each update of the generator, according to recommendations from previous research \cite{wGAN_paper} \cite{wGAN_GP_paper}. For each dataset (see Section \ref{Datasets}), a DVGAN is trained for 500 epochs on 6000 samples with a batch size of 512. A WGAN with the same configuration except for the inclusion of the derivative discriminator is also trained on each of the datasets to enable a comparison with DVGAN. The learning behaviour of the model can be monitored by plotting the losses of GAN components, which is observed in Figure \ref{fig:GAN_loss_comparison}. All models were designed using Python \textit{Keras} \cite{chollet2015keras} and \textit{Tensorflow} \cite{tensorflow2015-whitepaper} libraries\footnote{Computational resources were provided by the LIGO Laboratory and supported by the National Science Foundation Grants No. PHY-0757058 and No. PHY-0823459}.

It is important to note that effective GAN architectures are highly dependent on the data, and must be tuned according to the problem at hand. There is no optimization involved in either the WGAN or the DVGAN and there is no guarantee that the specified architectures are optimal in either case. Multiple model configurations were implemented that include different hyperparameters. Three such configurations are presented in Appendix \ref{DVGAN_architecture}, along with the final, most optimal model architecture (in bold text) found through trial and error (based on the evaluation followed in this paper). It is important to note that in many of the configurations attempted, the DVGAN offers more stable training (or at least similar) than WGAN counterparts. A uniform architecture is maintained across all experiments for a 1:1 comparison with the vanilla method. Thus, this research shows that it is possible to stabilize GAN training by bolstering the model with a second discriminator applied to another representation of the data.

\subsection{Training Data and Procedures}\label{Datasets}

\subsubsection{Dataset Construction and Preprocessing}

\begin{figure*}
\centering
\captionsetup{width = 0.85 \textwidth}
\includegraphics[width = 0.92\textwidth]{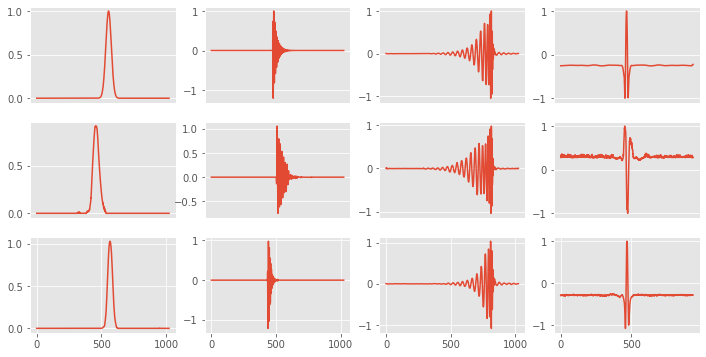}
\caption{Example waveforms (\textit{top}) for all datasets (pulse, ringdown, BBH, blip) with example generations from WGAN (\textit{middle}) and DVGAN (\textit{bottom}). Noise artifacts are more significant in the WGAN generations compared with DVGAN, in particular for the pulse and blip examples.}
\label{fig:Example_generations}
\end{figure*}

Four datasets comprising 1D differentiable signals are utilized to show the robustness of the DVGAN in this setting. The first three are inspired from \cite{McGinn_2021}, representing proxy waveforms analogous to GW waveforms and transients. The final dataset is compiled from actual `blip' glitch events and is obtained from \cite{GENGLI}.

\begin{enumerate}
    \item Gaussian pulses: short exponential increase then decrease in amplitude that follow the equation $h_{GP}(t) = exp(\frac{-t^2}{\tau^2})$. Gaussian pulses can represent anomalous glitch events observed in GW detectors.
    \item Ringdown:  damped oscillations mimicking the signals following a binary black hole collision, following the equation $h_{RD}(t) = Aexp[-(t-t_0)/\tau]sin(2\pi f_0(t-t_0)+\phi)$ with frequency $f_0$, duration parameter $\tau$, amplitude A, time of arrival $t_0$ and phase $\phi$ (uniformly sampled between [0, 2$\pi$]
    \item Binary Black Hole Coalescence (BBH): The inspiral and merger of a binary black hole system. All BBH signals were simulated using the IMRPhenom waveform routine from \textit{LALSuite} \cite{lalsuite}, which generates the inspiral, merger and ringdown of a BBH waveform. The component masses are restricted to the range of $[30, 70]M_{\odot}$ with a spin of zero and fixing $m_1>m_2$. The mass distribution is approximated by a power law with an index of 1.6 \cite{Abbott_2019}. The inclinations of the BBH signals are drawn so that the cosine of the angles lies uniformly in the range of $[-1, 1]$ using only plus polarization.
    \item Blip glitches: One glitch class of 22 that have been classified by \textit{Gravity Spy}, extracted and denoised from actual LIGO strain data. 
\end{enumerate}
    
The location of the peak amplitude of the Gaussian pulse and Ringdown datasets are randomly drawn from a uniform distribution within $[0.4, 0.6]$ from the start of the $1s$ time interval, with all training datasets except the blip glitch dataset sampled at $1024Hz$. This corresponds to signals of 1024 data points in length  for the first three datasets. GANs can generate higher dimensional spaces, however, larger networks and longer training times are generally required. Therefore, $1s$ segments are opted for with a sampling rate of $1024Hz$. The masses of the BBH dataset are restricted to be above 30 solar masses. Lower mass systems would produce longer-lasting waveforms that would fall outside the $1s$ interval defined. All training datasets are rescaled so that their peak amplitude is at 1. The Gaussian pulse and Ringdown datasets are analytic proxy waveforms of signals expected from burst GW sources. These two datasets and their suitability for this research are also described in \cite{Abbott_signal_description}. 

\vspace{-3mm}

The final blip glitch dataset was constructed using confidences from \textit{Gravity Spy} applied the blip glitch triggers in LIGO's second observing run (O2). In \cite{GENGLI}, they select blip glitch triggers from LIGO's Hanford (H1) and Livingston (L1) detectors with a high \textit{Gravity Spy} confidence, $c^1_{GS} \geq 0.9$, with $c^1_{GS}$ representing the base confidence provided by \textit{Gravity Spy} when applied to raw strain data. The glitches are extracted but are surrounded by stationary and uncorrelated noise, which would hinder the learning of GAN models. The dataset requires multiple preprocessing steps \cite{GENGLI}, which are also followed in this study, except for the final denoising step, called rROF \cite{Torres_2014_rrof}, which cannot remove some high-frequency artifacts that remain after previous preprocessing. Instead, to further reduce artifacts and assist in GAN training, two Savitzky-Golay smoothing filters \cite{savitzsky} are applied to the signals one after another (polynomials of order 3, windows of 21 and 11 respectively), which can smooth most noise artifacts in the dataset, and the smoothed blips are again rescaled. It is not investigated whether the characteristics of the blips are preserved after filtering. However, the signal shapes are analogous and can serve experimental purposes. With a sampling rate of $4096Hz$, the samples of the final training set have 938 data points, comprising $0.23s$ of LIGO strain data.

\section{Results}\label{Results}

\subsection{Data Generation and Experiments}
After training the WGAN and DVGAN on each training dataset, 6000 waveforms are generated by each network, examples of which can be viewed in Figure \ref{fig:Example_generations}. This is done by sampling a 100-dimensional vector from a normal distribution, typical of GAN studies. Three experiments are used to achieve a broad view of the performance on respective datasets: 

\begin{enumerate}
    \item \textit{Diversity} - A T-SNE analysis to ensure that the distribution of the generated signals aligns with those of the training set, showing that the diversity of the training set is captured.
    \item \textit{Fidelity} - A discriminative analysis that uses an off-the-shelf 2-layer, 1D CNN to classify a balanced dataset of real and generated samples (3000 samples each, trained over 20 epochs), giving a discriminative score, defined as $|0.5-HoldoutError|$.
    \item A final experiment, only applied to the BBH dataset, involves a match analysis on the generated samples, to investigate the parameter space covered by the generated signals compared to that of the training set. Metrics such as the mean maximum match, RMSE and the phase shift of the maximum match are used to show how well the generated data matches the training data, while the mass distributions are compared with that of the training data under a two-sample \textit{Kolmogorov–Smirnov} (KS) Test\cite{KSTest} at the 95\% confidence interval.
\end{enumerate}

The model architecture involved in the fidelity experiments is shown in Appendix \ref{Exp_networks}.

\begin{figure*}
\centering
    \setkeys{Gin}{width=0.24\textwidth}
{\includegraphics[scale=0.5]{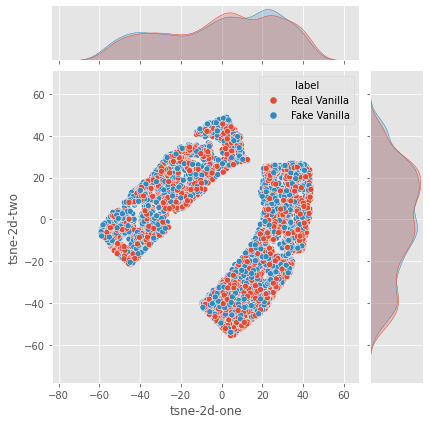}}
{\includegraphics[scale=0.5]{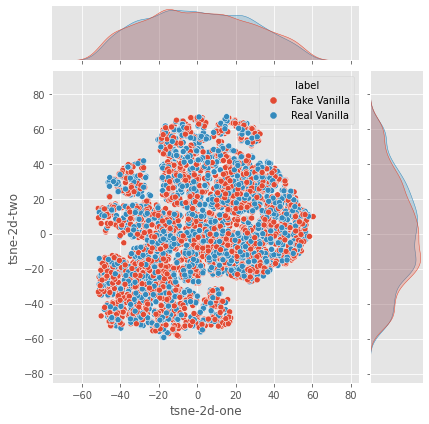}}
{\includegraphics[scale=0.5]{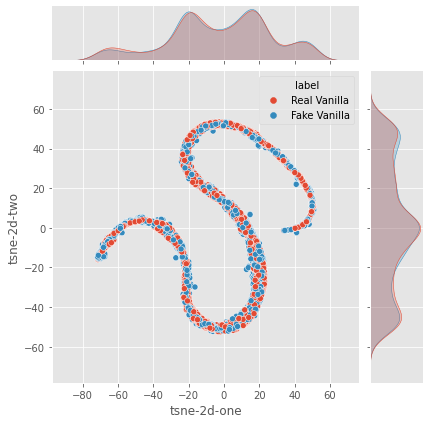}}
{\includegraphics[scale=0.5]{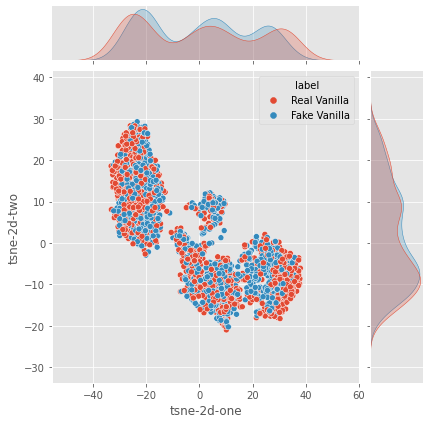}}
\end{figure*}
\vspace{0.00mm}
\begin{figure*}
\centering
\captionsetup{width = 0.85 \textwidth}
    \setkeys{Gin}{width=0.24\textwidth}
\subfloat[Pulse
          \label{fig:TSNE_pulse}]{\includegraphics[scale=0.5]{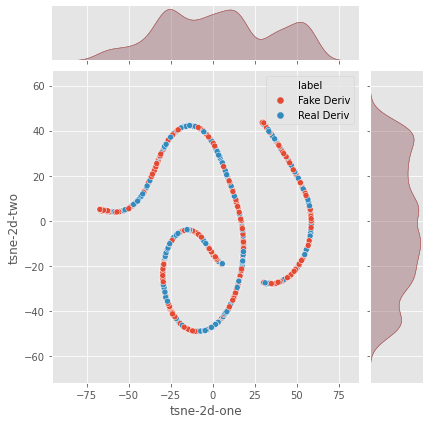}}
\subfloat[Ringdown
          \label{fig:TSNE_ringdown}]{\includegraphics[scale=0.5]{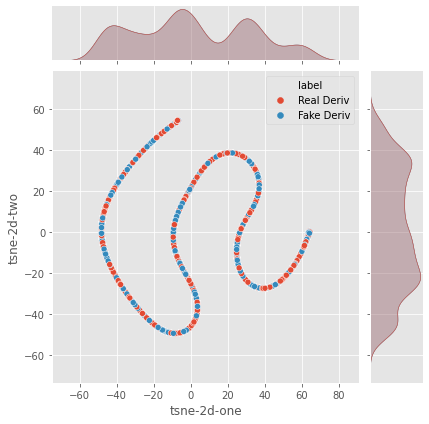}}
\subfloat[BBH
          \label{fig:TSNE_CBC}]{\includegraphics[scale=0.5]{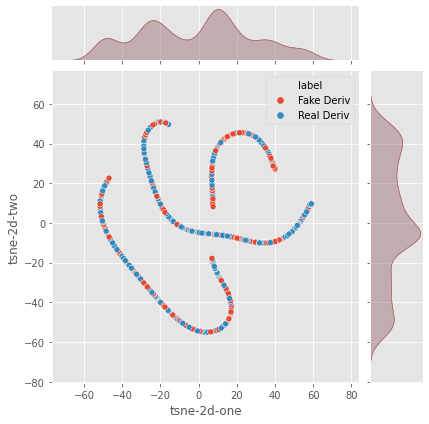}}
\subfloat[Blip
          \label{fig:TSNE_blip}]{\includegraphics[scale=0.5]{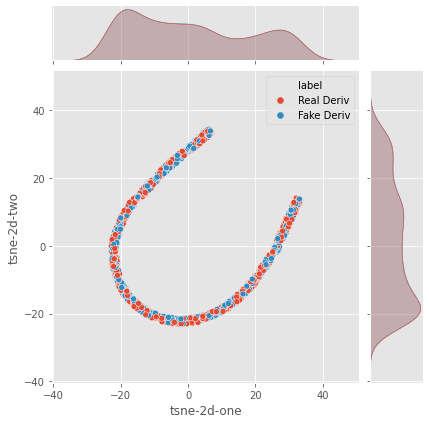}}

\caption{T-SNE plots comparing of real and synthetic signals for a vanilla WGAN (\textit{top}) and DVGAN (\textit{bottom})}
\label{fig:TSNE_plots}
\end{figure*}

\begin{table*}
\centering
\begin{tabular}{lll|lll} 
    \hline
    {} & {Discriminative} & {Score} & {} & {Match Analysis} \\ \hline \hline
\cline{3-3}
    {\textbf{Dataset}} & {\textbf{WGAN}} & \textbf{DVGAN} & {\textbf{Metric}} & {\textbf{WGAN}} & \textbf{DVGAN} \\ [0.1cm] \hline 
    Pulse & 0.029 $\pm$ $^{0.018}_{0.011}$ & \textbf{0.009} $\pm$ $^{0.008}_{0.008}$ & {Match } & \textbf{0.9918} $\pm$ $^{0.000}_{0.000}$ & 0.9917$\pm$ $^{0.000}_{0.000}$   \\ [0.2cm]
    Ringdown & 0.041 $\pm$ $^{0.029}_{0.009}$  & \textbf{0.017} $\pm$ $^{0.009}_{0.009}$  & RMSE & 0.0486 $\pm$ $^{0.0005}_{0.0007}$ & \textbf{0.0428} $\pm$ $^{0.0001}_{0.0002}$   \\ [0.2cm]
    BBH & 0.499 $\pm$ $^{0.000}_{0.001}$ & \textbf{0.199} $\pm$ $^{0.135}_{0.137}$ & Shift & 0.1740 $\pm$ $^{0.0019}_{0.0020}$ & \textbf{0.1535} $\pm$ $^{0.0015}_{0.0015}$\\ [0.2cm]
    Blip & 0.453 $\pm$ $^{0.012}_{0.010}$  & \textbf{0.129} $\pm$ $^{0.012}_{0.011}$ & {(KS1, KS2)} & (0.106 $\pm$ $^{0.007}_{0.010}$, 0.114 $\pm$ $^{0.014}_{0.008}$)    & (0.102 $\pm$ $^{0.011}_{0.021}$, 0.143 $\pm$ $^{0.009}_{0.005}$)  \\ [0.1cm]\hline \bottomrule
\end{tabular}
\captionsetup{width = 0.85 \textwidth}
\caption{Discriminative scores (lower is better) and match metrics for the WGAN and DVGAN on respective datasets. The results represent the mean over 5 iterations of experimentation. Bounds are provided by the maximum and minimum scores over 5 iterations. The Mean Max Match indicates, on average, how well the generated data matches training data signals, two KS test statistics measure the discrepancy between training data and estimated (synthetic) mass distributions (M1 and M2 respectively), and the shift describes how many data points signals must be shifted to maximize their match. Bounds are provided by the maximum and minimum scores over all iterations.}
\label{tab:discriminator_predictive_scores}
\end{table*}

\subsection{Experimental Analysis}

\subsubsection{Visualization - Diversity}

Visualizing the synthetic data under T-SNE allows for the examination of the diversity of the generated data compared with the training data. Figure \ref{fig:TSNE_plots} shows T-SNE plots comparing real and synthetic distributions in a reduced 2D space. It is observed that the generated data from both models cover the T-SNE space of the real distributions, yielding the diversity that is desired in generative models. However, it is observed that the generated data from DVGAN more closely matches that of the real data than that from WGAN. The distributions of the $x$ and $y$ axis support this. It is observed in the top plots of Figure \ref{fig:TSNE_plots} that there is a slight mismatch in distributions between real and synthetic, particularly for the x-axis. It is observed that the distributions of both T-SNE dimensions from DVGAN (Figure \ref{fig:TSNE_plots} bottom) match that of the real distribution almost exactly. This indicates that DVGAN can better match the real data while covering the corresponding distribution of data.

\subsubsection{Discriminative Score - Fidelity}
Table \ref{tab:discriminator_predictive_scores} shows the discriminative scores of WGAN against DVGAN across the respective datasets. The DVGAN yields significantly lower discriminative scores than the WGAN on all datasets, corresponding to more indistinguishable samples.

\subsubsection{Match Analysis for BBH Dataset}
Aside from comparing the generated data from GANs directly with the training data, it is also useful to compare the underlying parameters that result in such waveforms. For this purpose, a match analysis is conducted on the BBH waveforms from the WGAN and DGAN, using a large bank of 20,000 simulated BBH waveforms with known masses in the range $[20, 80]M_{\odot}$ (to investigate if the GANs generate waveforms with masses outside of the range in the training set). Both GANs are used to generate 500 waveforms for each iteration, which are compared with the bank of simulated BBHs using \textit{PyCBC's} \cite{PYCBC_cite} match function. Taking the masses of the maximum matches allows an investigation into the parameter space of the training data that is covered by the respective GANs. 

\begin{figure}[t!]
\centering
\captionsetup{width = 0.45 \textwidth}
\begin{subfigure}{0.25\textwidth}
  \centering
  \includegraphics[width =\textwidth]{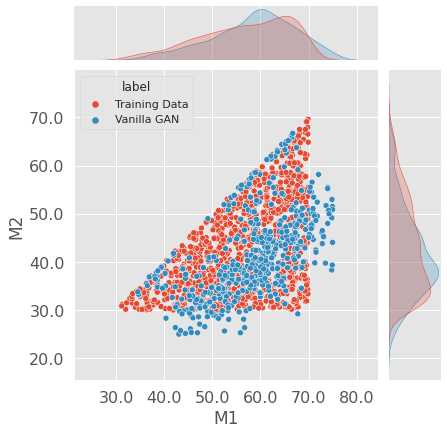}
  \caption{WGAN}
  \label{fig:Match_Filter_vanilla}
\end{subfigure}%
\begin{subfigure}{0.25\textwidth}
  \centering
  \includegraphics[width = \textwidth]{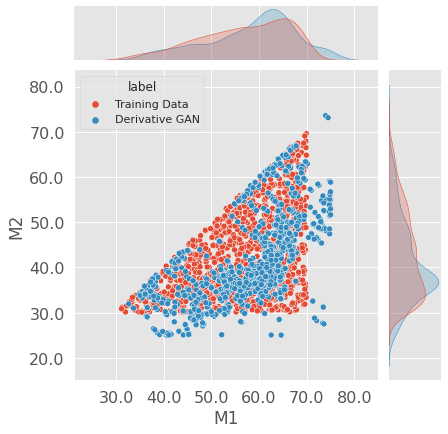}
  \caption{DVGAN}
  \label{fig:Match_Filter_deriv}
\end{subfigure}
\caption{Match Analysis on BBH generations from a Vanilla WGAN (\textit{left}) and DVGAN (\textit{right}).}
\label{fig:match_filter_vis}
\end{figure}

Figure \ref{fig:match_filter_vis} shows the coverage of the parameter space by the generated waveforms from respective models, compared with same-size subsets of the template bank. A lower triangular plot is observed in the template distribution, due to the fixing of $m_1>m_2$. For each GAN-generated waveform, the maximum match is recorded between it and its matched waveform to compare how well respective waveforms fit. Table \ref{tab:discriminator_predictive_scores} shows that a similar mean match is yielded by both models, however, the average RMSE from DVGAN generations is lower than that of WGAN, while the average shift is also lower. Since the phase of the BBH signals is kept constant, maximum matches should be achieved from a shift of zero. The lower shift value yielded by DVGAN generations implies that the generated data better matches the training data compared to WGAN. It is observed that the generated waveforms from both models have mass distributions that are different from that of the training data, and even match with waveforms outside of the mass range of the training data. This is confirmed by the two-sample KS tests, where the null hypothesis (same distribution) is rejected for each iteration of the experiment. However, it is unclear whether this is beneficial, given that interpolations in the signal space might correspond to extrapolations in the parameter space.

\section{Conclusion} \label{Conclusion}

\subsection{Conclusions}
This study has shown that the DVGAN outperforms the vanilla WGAN across multiple datasets and experiments, generating signals of higher fidelity. Specific model configurations have been employed to carry out an appropriate ablation study, although it has been found through experimentation that the DVGAN approach often yields smoother continuous signals of higher fidelity when compared to corresponding WGAN architectures, with minimal additional computational expense (if the auxiliary discriminator is kept small). Furthermore, the method has been tested on actual (filtered) blip glitch signals from LIGO, presenting the method's potential in this setting. The method might prove helpful in training data augmentation algorithms for GW physics, as well as other fields. Improvements upon techniques such as these could be useful for the training of fast, online reconstruction methods that can process huge quantities of data from multiple sources. 


\subsection{Future Work}
There are multiple avenues for further developments on this work. Rather than using an auxiliary discriminator applied to derivative signals, other representations of the data could be provided. For example, spectrogram representations could be provided to an auxiliary discriminator, or in the case of non-differentiable signals, a moving average or other relevant representations could be discriminated on. Higher-order derivatives could be included in some way in the second discriminator, possibly using padding to concatenate together. Then, 1D convolutions could be applied to the matrix. One problem typical of GANs is that they have trouble converging when trying to generate high-dimensional data, since the discriminator has too much information to distinguish real and fake samples, becoming too powerful too soon \cite{high_res_gan}. Caution would need to be taken in the case of including too much information in one discriminator so that it does not become too powerful. If computational resources are abundant, the auxiliary discriminator could be expanded to share a similar architecture as the base discriminator. This could further boost GAN training stability and performance. Only WGANs have been investigated in this study, and it would be interesting to understand the effectiveness under other configurations. It will be interesting to apply the DVGAN method to other GW signals, and in particular glitch signals, using \textit{Gravity Spy} confidences from the spectrogram data to test their fidelity. 

Implementing the method on other glitch classes might progress understanding of the GW detector glitch space, and augment datasets for classifiers. Conditioning on glitch classes and generating hybrid waveforms might enable classifiers to identify previously unknown GW signals.

    




\bibliographystyle{IEEEtran}
\bibliography{IEEEabrv,bibliography_DGAN}

\begin{thebibliography}{10}
\providecommand{\url}[1]{#1}
\csname url@samestyle\endcsname
\providecommand{\newblock}{\relax}
\providecommand{\bibinfo}[2]{#2}
\providecommand{\BIBentrySTDinterwordspacing}{\spaceskip=0pt\relax}
\providecommand{\BIBentryALTinterwordstretchfactor}{4}
\providecommand{\BIBentryALTinterwordspacing}{\spaceskip=\fontdimen2\font plus
\BIBentryALTinterwordstretchfactor\fontdimen3\font minus
  \fontdimen4\font\relax}
\providecommand{\BIBforeignlanguage}[2]{{%
\expandafter\ifx\csname l@#1\endcsname\relax
\typeout{** WARNING: IEEEtran.bst: No hyphenation pattern has been}%
\typeout{** loaded for the language `#1'. Using the pattern for}%
\typeout{** the default language instead.}%
\else
\language=\csname l@#1\endcsname
\fi
#2}}
\providecommand{\BIBdecl}{\relax}
\BIBdecl

\bibitem{medicine_GAN}
S.~Dash, A.~Yale, I.~Guyon, and K.~Bennett, \emph{Medical Time-Series Data
  Generation Using Generative Adversarial Networks}, 09 2020, pp. 382--391.

\bibitem{music_generation_survey}
\BIBentryALTinterwordspacing
S.~Ji, J.~Luo, and X.~Yang, ``A comprehensive survey on deep music generation:
  Multi-level representations, algorithms, evaluations, and future
  directions,'' 2020. [Online]. Available:
  \url{https://arxiv.org/abs/2011.06801}
\BIBentrySTDinterwordspacing

\bibitem{first_GW}
\BIBentryALTinterwordspacing
B.~P. Abbott~et al., ``Observation of gravitational waves from a binary black
  hole merger,'' \emph{Phys. Rev. Lett.}, vol. 116, p. 061102, Feb 2016.
  [Online]. Available:
  \url{https://link.aps.org/doi/10.1103/PhysRevLett.116.061102}
\BIBentrySTDinterwordspacing

\bibitem{CBC_detections_1}
\BIBentryALTinterwordspacing
B.~P. Abbott~et al, ``Gw151226: Observation of gravitational waves from a
  22-solar-mass binary black hole coalescence,'' \emph{Phys. Rev. Lett.}, vol.
  116, p. 241103, Jun 2016. [Online]. Available:
  \url{https://link.aps.org/doi/10.1103/PhysRevLett.116.241103}
\BIBentrySTDinterwordspacing

\bibitem{CBC_detections_3}
\BIBentryALTinterwordspacing
{The LIGO Scientific Collaboration et al.}, ``Gwtc-3: Compact binary
  coalescences observed by ligo and virgo during the second part of the third
  observing run,'' 2021. [Online]. Available:
  \url{https://arxiv.org/abs/2111.03606}
\BIBentrySTDinterwordspacing

\bibitem{Owen_1999_match}
\BIBentryALTinterwordspacing
B.~J. Owen and B.~S. Sathyaprakash, ``Matched filtering of gravitational waves
  from inspiraling compact binaries: Computational cost and template
  placement,'' \emph{Physical Review D}, vol.~60, no.~2, jun 1999. [Online].
  Available: \url{https://doi.org/10.1103%2Fphysrevd.60.022002}
\BIBentrySTDinterwordspacing

\bibitem{CCSNa}
C.~Fryer and K.~New, ``Gravitational waves from gravitational collapse,''
  \emph{Living Reviews in Relativity}, vol.~14, 12 2011.

\bibitem{Neutron_star}
L.~Baiotti, I.~Hawke, L.~Rezzolla, and E.~Schnetter, ``Details on the
  gravitational-wave emission from rotating gravitational collapse in 3d,''
  2007.

\bibitem{wave_burst}
\BIBentryALTinterwordspacing
S.~Klimenko, I.~Yakushin, A.~Mercer, and G.~Mitselmakher, ``A coherent method
  for detection of gravitational wave bursts,'' \emph{Classical and Quantum
  Gravity}, vol.~25, no.~11, p. 114029, may 2008. [Online]. Available:
  \url{https://doi.org/10.1088%2F0264-9381%2F25%2F11%2F114029}
\BIBentrySTDinterwordspacing

\bibitem{detchar_transient}
\BIBentryALTinterwordspacing
B.~P.~A. et~al., ``A guide to {LIGO}{\textendash}virgo detector noise and
  extraction of transient gravitational-wave signals,'' \emph{Classical and
  Quantum Gravity}, vol.~37, no.~5, p. 055002, feb 2020. [Online]. Available:
  \url{https://doi.org/10.1088%2F1361-6382%2Fab685e}
\BIBentrySTDinterwordspacing

\bibitem{ET_paper}
\BIBentryALTinterwordspacing
S.~H. et~al., ``Sensitivity studies for third-generation gravitational wave
  observatories,'' \emph{Classical and Quantum Gravity}, vol.~28, no.~9, p.
  094013, apr 2011. [Online]. Available:
  \url{https://doi.org/10.1088%2F0264-9381%2F28%2F9%2F094013}
\BIBentrySTDinterwordspacing

\bibitem{Cosmic_expl}
\BIBentryALTinterwordspacing
D.~Reitze~et al., ``Cosmic explorer: The u.s. contribution to
  gravitational-wave astronomy beyond ligo,'' 2019. [Online]. Available:
  \url{https://arxiv.org/abs/1907.04833}
\BIBentrySTDinterwordspacing

\bibitem{match_matched}
\BIBentryALTinterwordspacing
H.~Gabbard, M.~Williams, F.~Hayes, and C.~Messenger, ``Matching matched
  filtering with deep networks for gravitational-wave astronomy,'' \emph{Phys.
  Rev. Lett.}, vol. 120, p. 141103, Apr 2018. [Online]. Available:
  \url{https://link.aps.org/doi/10.1103/PhysRevLett.120.141103}
\BIBentrySTDinterwordspacing

\bibitem{GANpaper}
\BIBentryALTinterwordspacing
I.~J. Goodfellow~et al., ``Generative adversarial networks,'' 2014. [Online].
  Available: \url{https://arxiv.org/abs/1406.2661}
\BIBentrySTDinterwordspacing

\bibitem{TimeGANsSurvey}
\BIBentryALTinterwordspacing
E.~Brophy, Z.~Wang, Q.~She, and T.~Ward, ``Generative adversarial networks in
  time series: A survey and taxonomy,'' 2021. [Online]. Available:
  \url{https://arxiv.org/abs/2107.11098}
\BIBentrySTDinterwordspacing

\bibitem{GENGLI}
\BIBentryALTinterwordspacing
M.~Lopez, V.~Boudart, K.~Buijsman, A.~Reza, and S.~Caudill, ``Simulating
  transient noise bursts in ligo with generative adversarial networks,'' 2022.
  [Online]. Available: \url{https://arxiv.org/abs/2203.06494}
\BIBentrySTDinterwordspacing

\bibitem{GAN_spec_best}
\BIBentryALTinterwordspacing
J.~Yan, A.~P. Leung, and D.~C.~Y. Hui, ``On improving the performance of glitch
  classification for gravitational wave detection by using generative
  adversarial networks,'' 2022. [Online]. Available:
  \url{https://arxiv.org/abs/2207.04001}
\BIBentrySTDinterwordspacing

\bibitem{Omicron}
\BIBentryALTinterwordspacing
F.~Robinet, N.~Arnaud, N.~Leroy, A.~Lundgren, D.~Macleod, and J.~McIver,
  ``Omicron: A tool to characterize transient noise in gravitational-wave
  detectors,'' \emph{SoftwareX}, vol.~12, p. 100620, 2020. [Online]. Available:
  \url{https://www.sciencedirect.com/science/article/pii/S2352711020303332}
\BIBentrySTDinterwordspacing

\bibitem{RCGANpaper}
\BIBentryALTinterwordspacing
C.~Esteban, S.~L. Hyland, and G.~Rätsch, ``Real-valued (medical) time series
  generation with recurrent conditional gans,'' 2017. [Online]. Available:
  \url{https://arxiv.org/abs/1706.02633}
\BIBentrySTDinterwordspacing

\bibitem{TimeGANpaper}
\BIBentryALTinterwordspacing
J.~Yoon, D.~Jarrett, and M.~van~der Schaar, ``Time-series generative
  adversarial networks,'' in \emph{Advances in Neural Information Processing
  Systems}, H.~Wallach, H.~Larochelle, A.~Beygelzimer, F.~d\textquotesingle
  Alch\'{e}-Buc, E.~Fox, and R.~Garnett, Eds., vol.~32.\hskip 1em plus 0.5em
  minus 0.4em\relax Curran Associates, Inc., 2019. [Online]. Available:
  \url{https://proceedings.neurips.cc/paper/2019/file/c9efe5f26cd17ba6216bbe2a7d26d490-Paper.pdf}
\BIBentrySTDinterwordspacing

\bibitem{sigWasserstein}
\BIBentryALTinterwordspacing
H.~Ni, L.~Szpruch, M.~Wiese, S.~Liao, and B.~Xiao, ``Conditional
  sig-wasserstein gans for time series generation,'' 2020. [Online]. Available:
  \url{https://arxiv.org/abs/2006.05421}
\BIBentrySTDinterwordspacing

\bibitem{McGinn_2021}
\BIBentryALTinterwordspacing
J.~McGinn, C.~Messenger, M.~J. Williams, and I.~S. Heng, ``Generalised
  gravitational wave burst generation with generative adversarial networks,''
  \emph{Classical and Quantum Gravity}, vol.~38, no.~15, p. 155005, jun 2021.
  [Online]. Available: \url{https://doi.org/10.1088\%2F1361-6382\%2Fac09cc}
\BIBentrySTDinterwordspacing

\bibitem{GENGLI_2}
\BIBentryALTinterwordspacing
M.~Lopez, V.~Boudart, S.~Schmidt, and S.~Caudill, ``Simulating transient noise
  bursts in ligo with gengli,'' 2022. [Online]. Available:
  \url{https://arxiv.org/abs/2205.09204}
\BIBentrySTDinterwordspacing

\bibitem{Zevin_2017}
\BIBentryALTinterwordspacing
M.~Z. et~al., ``Gravity spy: integrating advanced {LIGO} detector
  characterization, machine learning, and citizen science,'' \emph{Classical
  and Quantum Gravity}, vol.~34, no.~6, p. 064003, feb 2017. [Online].
  Available: \url{https://doi.org/10.1088\%2F1361-6382\%2Faa5cea}
\BIBentrySTDinterwordspacing

\bibitem{gspy_data_quality}
\BIBentryALTinterwordspacing
J.~Glanzer and et~al, ``Data quality up to the third observing run of advanced
  ligo: Gravity spy glitch classifications,'' 2022. [Online]. Available:
  \url{https://arxiv.org/abs/2208.12849}
\BIBentrySTDinterwordspacing

\bibitem{3playerGANhard}
\BIBentryALTinterwordspacing
S.~Vandenhende, B.~De~Brabandere, D.~Neven, and L.~Van~Gool, ``A three-player
  gan: Generating hard samples to improve classification networks,'' 2019.
  [Online]. Available: \url{https://arxiv.org/abs/1903.03496}
\BIBentrySTDinterwordspacing

\bibitem{3playerface}
Y.~Shen, P.~Luo, P.~Luo, J.~Yan, X.~Wang, and X.~Tang, ``Faceid-gan: Learning a
  symmetry three-player gan for identity-preserving face synthesis,'' in
  \emph{2018 IEEE/CVF Conference on Computer Vision and Pattern Recognition},
  2018, pp. 821--830.

\bibitem{3playerdual}
\BIBentryALTinterwordspacing
T.~D. Nguyen, T.~Le, H.~Vu, and D.~Phung, ``Dual discriminator generative
  adversarial nets,'' 2017. [Online]. Available:
  \url{https://arxiv.org/abs/1709.03831}
\BIBentrySTDinterwordspacing

\bibitem{wGAN_paper}
\BIBentryALTinterwordspacing
M.~Arjovsky, S.~Chintala, and L.~Bottou, ``{W}asserstein generative adversarial
  networks,'' in \emph{Proceedings of the 34th International Conference on
  Machine Learning}, ser. Proceedings of Machine Learning Research, D.~Precup
  and Y.~W. Teh, Eds., vol.~70.\hskip 1em plus 0.5em minus 0.4em\relax PMLR,
  06--11 Aug 2017, pp. 214--223. [Online]. Available:
  \url{https://proceedings.mlr.press/v70/arjovsky17a.html}
\BIBentrySTDinterwordspacing

\bibitem{wGAN_GP_paper}
\BIBentryALTinterwordspacing
I.~Gulrajani, F.~Ahmed, M.~Arjovsky, V.~Dumoulin, and A.~Courville, ``Improved
  training of wasserstein gans,'' 2017. [Online]. Available:
  \url{https://arxiv.org/abs/1704.00028}
\BIBentrySTDinterwordspacing

\bibitem{karras2018progressive}
\BIBentryALTinterwordspacing
T.~Karras, T.~Aila, S.~Laine, and J.~Lehtinen, ``Progressive growing of {GAN}s
  for improved quality, stability, and variation,'' in \emph{International
  Conference on Learning Representations}, 2018. [Online]. Available:
  \url{https://openreview.net/forum?id=Hk99zCeAb}
\BIBentrySTDinterwordspacing

\bibitem{improved_GAN_training}
\BIBentryALTinterwordspacing
T.~Salimans, I.~Goodfellow, W.~Zaremba, V.~Cheung, A.~Radford, and X.~Chen,
  ``Improved techniques for training gans,'' 2016. [Online]. Available:
  \url{https://arxiv.org/abs/1606.03498}
\BIBentrySTDinterwordspacing

\bibitem{improving_improved_training}
\BIBentryALTinterwordspacing
X.~Wei, B.~Gong, Z.~Liu, W.~Lu, and L.~Wang, ``Improving the improved training
  of wasserstein gans: A consistency term and its dual effect,'' 2018.
  [Online]. Available: \url{https://arxiv.org/abs/1803.01541}
\BIBentrySTDinterwordspacing

\bibitem{Original_wasserstein}
\BIBentryALTinterwordspacing
M.~Arjovsky, S.~Chintala, and L.~Bottou, ``Wasserstein gan,'' 2017. [Online].
  Available: \url{https://arxiv.org/abs/1701.07875}
\BIBentrySTDinterwordspacing

\bibitem{discriminator_training}
\BIBentryALTinterwordspacing
N.~Kodali, J.~Abernethy, J.~Hays, and Z.~Kira, ``On convergence and stability
  of gans,'' 2017. [Online]. Available: \url{https://arxiv.org/abs/1705.07215}
\BIBentrySTDinterwordspacing

\bibitem{chollet2015keras}
\BIBentryALTinterwordspacing
F.~Chollet \emph{et~al.} (2015) Keras. [Online]. Available:
  \url{https://github.com/fchollet/keras}
\BIBentrySTDinterwordspacing

\bibitem{tensorflow2015-whitepaper}
\BIBentryALTinterwordspacing
M.~A. et~al., ``{TensorFlow}: Large-scale machine learning on heterogeneous
  systems,'' 2015, software available from tensorflow.org. [Online]. Available:
  \url{http://tensorflow.org/}
\BIBentrySTDinterwordspacing

\bibitem{lalsuite}
{LIGO Scientific Collaboration}, ``{LIGO} {A}lgorithm {L}ibrary - {LALS}uite,''
  free software (GPL), 2018.

\bibitem{Abbott_2019}
\BIBentryALTinterwordspacing
B.~P.~A. et~al., ``Binary black hole population properties inferred from the
  first and second observing runs of advanced {LIGO} and advanced virgo,''
  \emph{The Astrophysical Journal}, vol. 882, no.~2, p. L24, sep 2019.
  [Online]. Available: \url{https://doi.org/10.3847/2041-8213/ab3800}
\BIBentrySTDinterwordspacing

\bibitem{Abbott_signal_description}
\BIBentryALTinterwordspacing
B.~P. Abbott~et al., ``All-sky search for short gravitational-wave bursts in
  the second advanced ligo and advanced virgo run,'' \emph{Phys. Rev. D}, vol.
  100, p. 024017, Jul 2019. [Online]. Available:
  \url{https://link.aps.org/doi/10.1103/PhysRevD.100.024017}
\BIBentrySTDinterwordspacing

\bibitem{Torres_2014_rrof}
\BIBentryALTinterwordspacing
A.~Torres, A.~Marquina, J.~A. Font, and J.~M. Ib{\'{a}}{\~{n}}ez,
  ``Total-variation-based methods for gravitational wave denoising,''
  \emph{Physical Review D}, vol.~90, no.~8, oct 2014. [Online]. Available:
  \url{https://doi.org/10.1103%2Fphysrevd.90.084029}
\BIBentrySTDinterwordspacing

\bibitem{savitzsky}
N.~Gallagher, ``Savitzky-golay smoothing and differentiation filter,'' 01 2020.

\bibitem{KSTest}
\BIBentryALTinterwordspacing
\emph{Kolmogorov--Smirnov Test}.\hskip 1em plus 0.5em minus 0.4em\relax New
  York, NY: Springer New York, 2008, pp. 283--287. [Online]. Available:
  \url{https://doi.org/10.1007/978-0-387-32833-1_214}
\BIBentrySTDinterwordspacing

\bibitem{PYCBC_cite}
\BIBentryALTinterwordspacing
A.~N. et~al., ``gwastro/pycbc: v2.0.4 release of pycbc,'' Jun. 2022. [Online].
  Available: \url{https://doi.org/10.5281/zenodo.6646669}
\BIBentrySTDinterwordspacing

\bibitem{high_res_gan}
\BIBentryALTinterwordspacing
T.~Miyato, T.~Kataoka, M.~Koyama, and Y.~Yoshida, ``Spectral normalization for
  generative adversarial networks,'' 2018. [Online]. Available:
  \url{https://arxiv.org/abs/1802.05957}
\BIBentrySTDinterwordspacing

\end{thebibliography}

\newpage
\onecolumn
\appendix
\vspace{-3mm}
\subsection{DVGAN Architecture} \label{DVGAN_architecture}
\vspace{-3mm}
\begin{table}[h]
\centering
\begin{tabular}{llllll} 
    \toprule
    {} & {} & {Discriminator} & {} & {} & {(2.5M, 3.5M, \textbf{4.1M} param.)} \\ \hline
    {Operation} & {Output shape} & {Kernel size} & {Stride} & {Dropout} & {Activation} \\ \hline
    {Signal input} & (1024) & {-} & {-}  & {0}  & {-}\\ 
    {Reshape} & (1024,1) & {-} & {-}  & {0}  & {-}\\ 
    {Convolutional} & (512,32), (-), \textbf{(-)} & {5} & {2}  & {0.5}  & {Leaky ReLU}\\ 
    {Convolutional} & (256,64), (-), \textbf{(512,64)} & {5} & {2}  & {0.5}  & {Leaky ReLU}\\ 
    {Convolutional} & (128, 128), (512,64), \textbf{(256,128)} & {5} & {2}  & {0.5}  & {Leaky ReLU}\\ 
    {Convolutional} & (64, 256), (256,128), \textbf{(128,256)} & {5} & {2}  & {0.5}  & {Leaky ReLU}\\ 
    {Convolutional} & (32, 512), (128,256), \textbf{(64,512)} & {5} & {2}  & {0.5}  & {Leaky ReLU}\\ 
    {Flatten} & (16384), (32768), \textbf{(32768)} & {-} & {-}  & {0.5}  & {-}\\ 
    {Dense} & (100) & {-} & {-}  & {0.2}  & {Leaky ReLU}\\ 
    {Dense} & (1) & {-} & {-}  & {0}  & {Sigmoid}\\ \bottomrule
    
    {} & {} & {DV Discriminator} & {} & {} & {(300k param.)} \\ \hline
    {Operation} & {Output shape} & {Kernel size} & {Stride} & {Dropout} & {Activation} \\ \hline
    {Signal input} & (1023) & {-} & {-}  & {0}  & {-}\\ 
    {Dense} & (1024) & {-} & {-}  & {0}  & {-}\\ 
    {Reshape} & (512,2) & {-} & {-}  & {0}  & {-}\\ 
    {Convolutional} & (256,64) & {5} & {2}  & {0.5}  & {Leaky ReLU}\\ 
    {Convolutional} & (128,128) & {5} & {2}  & {0.5}  & {Leaky ReLU}\\ 
    {Convolutional} & (64,256) & {5} & {2}  & {0.5}  & {Leaky ReLU}\\ 
    {Flatten} & (16384) & {-} & {-}  & {0.5}  & {-}\\ 
    {Dense} & (1) & {-} & {-}  & {0}  & {Sigmoid}\\ \bottomrule
    
    {} & {} & {Generator} & {} & {} & {(2.5M, 3.5M, \textbf{4.1M} param.)} \\ \hline
    {Operation} & {Output shape} & {Kernel size} & {Stride} & {BN} & {Activation} \\ \hline
    {Latent input} & (100) & {-} & {-}  & {\xmark}  & {-}\\ 
    {Dense} & (16384), (32768), \textbf{(32768)} & {-} & {-}  & {\xmark}  & {ReLU}\\ 
    {Reshape} & (32,512), (128,256), \textbf{(64,512)} & {-} & {-}  & {\xmark}  & {-}\\ 
    {Transposed conv.} & (64,256), (-), \textbf{(-)} & {5} & {2}  & {\cmark}  & {ReLU}\\ 
    {Transposed conv.} & (128,128), (-), \textbf{(128,256)} & {5} & {2}  & {\cmark}  & {ReLU}\\ 
    {Transposed conv.} & (256,64), (256,128), \textbf{(256,128)} & {5} & {2}  & {\xmark}  & {ReLU}\\ 
    {Transposed conv.} & (512,32), (512,64), \textbf{(512,64)} & {5} & {2}  & {\xmark}  & {ReLU}\\ 
    {Transposed conv.} & (1024,1), (1024, 1), \textbf{(1024,1)} & {5} & {2}  & {\xmark}  & {Linear}\\ 
    {Reshape} & (1024) & {-} & {-}  & {\xmark}  & {-} \\ \hline
    {Optimizer} & {RMSprop($\alpha = 0.0001$)} & {} & {} & {} & {} \\
    {Batch size} & {512} & {} & {} & {} & {} \\
    {Epochs} & {500} & {} & {} & {} & {} \\
    {Loss} & {Wasserstein} & {} & {} & {} & {} \\ \bottomrule
\end{tabular}
\captionsetup{width = 0.85 \textwidth}
\caption{The architecture and hyperparameters describing DVGAN, which consists of a base discriminator, a derivative (DV) discriminator and generator convolutional networks. The base discriminator and generator of the final configuration (bold text) each consist of four convolutional/transposed convolutional layers. The architecture is modified slightly for the blip glitch dataset, given a different input shape. In this case, a Dense layer of 938 neurons (LeakyReLU and ReLU activations respectively) is added before convolutions in the discriminator and after transposed convolutions in the generator. The table shows the hyperparameters of three model configurations that include a different number of layers and neurons, shown in brackets in the output shape and the top of each sub-table. Bold text indicates the optimal configuration that was used for the results presented. It is important to note that the WGANs implemented in this study follow the same architecture, only omitting the DV discriminator.}
\label{tab:model_architecture}
\end{table}

\subsection{CNN Network for Discriminative Score} \label{Exp_networks}

\begin{table}[H]
\centering
\begin{tabular}{llllll} 
    \toprule
    {Operation} & {Output shape} & {Kernel size} & {Stride} & {Dropout} & {Activation} \\ \hline
    {Signal input} & (1024) & {-} & {-}  & {0}  & {-}\\ 
    {Reshape} & (1024,1) & {-} & {-}  & {0}  & {-}\\ 
    {Convolutional} & (512,32) & {5} & {2}  & {0.5}  & {Leaky ReLU}\\ 
    {Convolutional} & (256,64) & {5} & {2}  & {0.5}  & {Leaky ReLU}\\ 
    {Flatten} & (16384) & {-} & {-}  & {0.5}  & {-}\\ 
    {Dense} & (1) & {-} & {-}  & {0}  & {Sigmoid}\\ \bottomrule
    
    {Optimizer} & {RMSprop($\alpha = 0.0001$)} & {} & {} & {} & {} \\
    {Batch size} & {64} & {} & {} & {} & {} \\
    {Epochs} & {20} & {} & {} & {} & {} \\
    {Loss} & {Binary Crossentropy} & {} & {} & {} & {} \\ \bottomrule
\end{tabular}
\captionsetup{width = 0.85 \textwidth}
\caption{Post-hoc Discriminative CNN (27k params.). The CNN applied to blips is the same except for the input signal shape.}
\label{tab:CNN_model_architecture}
\end{table}

\end{document}